\documentclass[
 nofootinbib,
 amsmath,amssymb,
 aps, showpacs, superscriptaddress
]{revtex4}

\usepackage{mfirstuc} 
\newcommand{\addReviewer}[2]{
  \expandafter\newcommand\csname #1\endcsname[1]{{\bf \color{#2} \capitalisewords{#1}:\,##1}}
  \expandafter\newcommand\csname #1cor\endcsname[2]{{\color{#2} \capitalisewords{#1}:\,\st{##1}{\bf ##2}}}
  \expandafter\newcommand\csname #1color\endcsname{#2}
}

\addReviewer{enrico}{blue}
\addReviewer{nathan}{purple}

\usepackage{feynmp-auto}

\usepackage{scalerel}

\newcommand{\overbar}[1]{\mkern 1.5mu\overline{\mkern-1.5mu#1\mkern-1.5mu}\mkern 1.5mu}

\usepackage{epstopdf}
\usepackage{siunitx} 
\usepackage{xparse}
\usepackage{fancyhdr}
\usepackage{soul}
\usepackage{comment}
\NewDocumentCommand{\fr}{>{\SplitArgument{1}{,}}m}{\efrac#1}
\NewDocumentCommand{\efrac}{mm}{\ensuremath{\frac{#1}{#2}}}

\usepackage{graphicx}
\usepackage[final]{pdfpages} 
\usepackage{dcolumn}
\usepackage{bm}
\usepackage{amssymb,braket,slashed,enumerate,bbold,empheq,xcolor,cases}

\newcommand{\ti}{\textit}
\newcommand{\tb}{\textbf}

\DeclareMathOperator{\eps}{\epsilon}

\usepackage[colorlinks=true,
            linkcolor=blue,
            urlcolor=blue,
            citecolor=blue]{hyperref}

\usepackage[noabbrev]{cleveref}

\begin{document}

\preprint{APS/123-QED}

\title{Lorentz violation and the electron-ion collider}

\author{Enrico\ Lunghi}
\affiliation{Physics Department, Indiana University, 
Bloomington, IN 47405, USA
}
\affiliation{Indiana University Center for Spacetime Symmetries, 
Bloomington, IN 47405, USA
}
\affiliation{Center for Exploration of Energy and Matter,
Indiana University, Bloomington, IN 47403, USA}
\affiliation{INFN Sezione di Torino}
\author{Nathan Sherrill}
\affiliation{Physics Department, Indiana University, 
Bloomington, IN 47405, USA
}
\affiliation{Indiana University Center for Spacetime Symmetries, 
Bloomington, IN 47405, USA
}
\affiliation{Center for Exploration of Energy and Matter,
Indiana University, Bloomington, IN 47403, USA}


\date{\today}

\begin{abstract}
We investigate the prospects for detecting violations of Lorentz symmetry in unpolarized deep inelastic electron-proton scattering in the context
of the future electron-ion collider. Simulated differential cross-section data are used to place expected bounds on a class of quark-sector coefficients for Lorentz violation that induce sidereal time dependence in the scattering cross section. We find that, with $100 \; {\rm fb}^{-1}$ of integrated luminosity, the expected bounds are in the $10^{-5}-10^{-7}$ range and are roughly two orders of magnitude stronger than those that can be extracted from existing HERA data. We also discuss the possibility of extracting bounds on the remaining time-independent coefficients.
\end{abstract}

  \maketitle
\section{\label{sec:intro}Introduction} 

     Lorentz invariance is a global symmetry of the Standard Model (SM) of particle physics and a local symmetry of General Relativity. While both theories have been fantastically successful in describing physics at currently attainable energies, it is widely expected that a fully quantum-theoretical description of all known physics including gravity will emerge at the Planck scale. One interesting possible consequence of this unification is the violation of Lorentz invariance. It was first shown by Kosteleck\'y and Samuel in Ref.~\cite{string} that the mechanism of spontaneous symmetry breaking could generate Lorentz violation in string theory. In this setting, the low-energy limit of this theory gains terms in its Lagrange density that take the general form \cite{stringscpt}
\begin{equation}
\label{eq:LVgeneral}
\mathcal{L}_{\text{LV}} \sim \frac{\lambda}{m_{\text{P}}^{k}}\braket{T}\cdot\bar{\psi}\Gamma(i\partial)^{k}\chi + \text{h.c.} ,
\end{equation}
where $\lambda$ is a dimensionless coupling constant, $k$ is an integer exponent, and $m_{\text{P}}$ is the Planck mass. The object $\braket{T}$ is a nonzero vacuum expectation value (vev) of a tensor field with suppressed spacetime indices, and $\Gamma$ is a generic gamma-matrix structure. The fields $\psi, \chi$ are generic four-dimensional fermion fields. In Eq.~\eqref{eq:LVgeneral}, Lorentz symmetry is spontaneously broken by the vev $\langle T \rangle$, which has orientation dependence (i.e. it not a scalar). Note that the underlying theory is Poincar\'e invariant, thus preserving microcausality, the spin-statistics theorem, the positivity of energy, power-counting renormalizability, standard quantization, and observer Lorentz invariance. Moreover, if $\braket{T}$ is a spacetime constant, energy-momentum conservation is also preserved.

To date, all high-precision tests of the Lorentz symmetry in the SM and gravity give no indication of Lorentz violation. Nevertheless, as we explained above, it is reasonable to entertain the possibility that Lorentz invariance is spontaneously broken at Planckian scales. The huge gap between these scales and those currently accessible at colliders (roughly 15 orders of magnitude) makes it impossible to detect directly the degrees of freedom responsible for the potential breaking of Lorentz symmetry.
An alternative approach is to search for suppressed signals at attainable energies. Probing Nature in this way suggests the use of a low-energy, effective quantum field theory which completely accounts for all possible residual Lorentz-violating effects that presumably originate from mechanisms in a more fundamental theory. This framework exists and is known as the Standard-Model Extension (SME) \cite{SME1,SME2,SME3}. For some accessible reviews of the SME, we refer the reader to Refs.~\cite{SMEoverview1,SMEoverview2} and references therein. By construction, the SME contains the field content from all known fundamental physics with the addition of all possible terms built from fundamental fields that break Lorentz and CPT symmetry. 
\noindent These additional terms take the form of coefficients contracted with products of SM and gravitational fields. As an example, consider the quantum-electrodynamics (QED) extension of the SM \cite{SME2}:
\begin{align}
\label{eq:LVQED}
\mathcal{L}_{\text{QED}}^{\text{ext.}} = &\tfrac{1}{2}i\bar{\psi}\Gamma^{\nu}\overset{\text{\tiny$\leftrightarrow$}}D_{\nu}\psi - \bar{\psi}M\psi -\tfrac{1}{4}F_{\mu\nu}F^{\mu\nu} \nonumber\\
& -\tfrac{1}{4}\left(\kappa_{F}\right)_{\kappa\lambda\mu\nu}F^{\kappa\lambda}F^{\mu\nu} + \tfrac{1}{2}\left(\kappa_{AF}\right)^{\kappa}\epsilon_{\kappa\lambda\mu\nu}A^{\lambda}F^{\mu\nu},
\end{align}
where
\begin{align}
& \Gamma^{\nu} = \gamma^{\nu} + c^{\mu\nu}\gamma_{\mu} + d^{\mu\nu}\gamma_{5}\gamma_{\mu} + e^{\nu} + if^{\nu}
\gamma_{5} + \tfrac{1}{2}g^{\lambda\mu\nu}\sigma_{\lambda\mu}, \nonumber\\
& M = m + a_{\mu}\gamma^{\mu} + b_{\mu}\gamma_{5}\gamma^{\mu} + \tfrac{1}{2}H_{\mu\nu}\sigma^{\mu\nu}.
\end{align}
Here, the coefficients for CPT and Lorentz violation are $a^\mu, b^\mu, c^{\mu\nu}, d^{\mu\nu}, e^{\nu}, f^{\nu}, g^{\lambda\mu\nu}, H^{\mu\nu}$ in the fermion sector and $\left(\kappa_{AF}\right)^{\kappa}, \left(\kappa_{F}\right)_{\kappa\lambda\mu\nu}$ in the photon sector. These coefficients are real quantities that can be thought of as a coupling constants or vevs (see Eq.~\eqref{eq:LVgeneral}).
An important property of the coefficients for Lorentz violation is that they transform as tensors under general coordinate transformations, called Lorentz observer transformations, but as scalars under transformations of the physical system itself, called Lorentz particle transformations \cite{SME1}. Because these coefficients represent preferred directions in spacetime, their presence implies a violation of Lorentz symmetry. For local quantum field theories, CPT symmetry is related to Lorentz symmetry through the CPT theorem \cite{CPT}. This means that CPT-violating effects 
are also completely parametrized by the SME. Thus, the SME can be understood as a general phenomenological framework used to search for CPT- and Lorentz-violating suppressed signals arising from a more fundamental theory. 
We remind the reader that the SME parametrizes all possible ways Lorentz and CPT symmetry can be violated in terms of known physical fields under the assumption of preserved locality and hermiticity. Therefore, in light of Eq.~\eqref{eq:LVgeneral}, there are in principle an infinite number of CPT- and Lorentz-breaking operators with an increasing number of derivatives. At energies well below a prescribed high-energy scale (e.g., the Planck mass), it is sufficient to restrict attention to operators of mass dimension four or less so that power-counting renormalizability and gauge invariance are satisfied.  This subset of the SME is referred to as the minimal SME (mSME). 
The mSME thus has all of the properties of the usual SM except that Lorentz invariance is broken by particle Lorentz transformations, and CPT is violated in the presence of CPT-odd operators---see Ref.~\cite{SME1,SME2} for a complete listing of all of terms appearing in the mSME.

Constraints on many coefficients for Lorentz violation across all sectors of the SME have been placed to date \cite{datatables}. Despite the large amount of work that has been carried out thus far, comparatively little attention has been paid to the quantum-chromodynamics (QCD) sector of the SME. This is primarily due to the difficulties in bypassing the observed spectrum of states to access the fundamental degrees of freedom of QCD. Very recently, there has been a push towards exploring Lorentz violation in this sector \cite{topquark,hadronic1,hadronic2,hadronic3,hadronic4,LVdis}. Much of this work may ultimately be relevant to the proposed electron-ion collider (EIC) \cite{EICsummary}, which is expected to usher in a new era of precision QCD studies of the hadrons and nuclei. The collider itself expected to be constructed at either the Thomas Jefferson National Laboratory (JLab) or Brookhaven National Laboratory (BNL). Once built, it will be the only collider capable of controlling the polarization of both the lepton and ion beams, which will enable an unprecedented understanding of the nucleon's spin content and tomography. Current design parameters for the JLab EIC (JLEIC) and BNL EIC (eRHIC) suggest a similar reach in terms of kinematical phase space \cite{JLEICdesign,eRHICdesign}. In this regard, the main distinction between the two currently proposed designs is that the JLEIC is expected to have a lower center of mass (CM) energy range than the eRHIC, but a higher luminosity. Whether the EIC is built at JLab or BNL, each design will be capable of being upgraded to a comparable CM energy and luminosity. Thus, in principle the only distinction between the two proposed designs is their geographic location and colliding beamline orientations. In the context of Lorentz violation, these traits become relevant. In this work we explore some of the consequences of these facts. Since the EIC will have a unique ability to study QCD, it is interesting to consider the prospects for detecting effects emanating from Lorentz-violating QCD. This is the basis for the current document, which examines the prospects for detecting Lorentz violation at the EIC through the process of unpolarized electron-proton deep inelastic scattering ($e$-$p$ DIS). 

\section{Lorentz-violating effects in unpolarized DIS}\label{sec:phenom}
 \subsection{General Setup for Unpolarized DIS}
\begin{figure}[t]
\centering
\includegraphics[width=0.4 \linewidth]{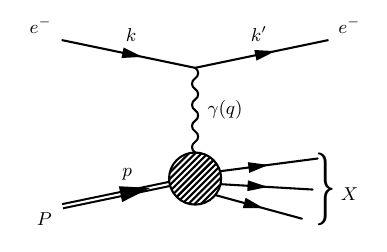}
\caption{Inclusive $e$-$p$ DIS in the one-photon exchange approximation. An electron $e^{-}$ with momentum $k$ interacts and recoils with momentum $k'$ from a proton $P$ with momentum $p$ through the exchange of a photon $\gamma$ producing a generic final hadronic state $X$. \label{fig:DISgeneral}}
\end{figure}

We now turn our attention to the process of inclusive $e$-$p$ DIS, which is illustrated in Fig.~\ref{fig:DISgeneral}. 
The observable of interest is the differential cross section. This can be written as
\begin{equation}
\label{eq:diffxsec}
d\sigma = \sum_{X}\int d\Pi_{X}(2\pi)^{4}\delta^{4}\left(p + k - k' - p_{X}\right)\fr{d^{3}k',(2\pi)^{3}2E'}\fr{|\mathcal{M}|^{2},F},
\end{equation}
where $\mathcal{M}$ denotes the spin-averaged scattering amplitude, and $F$ is the flux factor for the colliding particles. The rest of the expression is the phase-space element, which includes a sum over all possible (unobserved) hadronic states $X$ carrying a net momentum $p_{X}$ and the outgoing final-state electron density of states. We first focus on the properties of the amplitude $\mathcal{M}$.
For the DIS process depicted in Fig.~\ref{fig:DISgeneral}, the amplitude consists of the lepton vertex, intermediate propagator, and hadronic vertex. The structure of the former two are well understood. The hadronic contribution to the scattering amplitude takes the form of a hadronic neutral current $J_{\mu}(x)$ inserted between the initial proton $\ket{P}$ and final hadronic $\ket{X}$ state: 
\begin{equation}
\label{eq:hadronicamp}
\mathcal{M}_{\text{had.}} \sim \eps^{\mu}\bra{X}J_{\mu}(0)\ket{P},
\end{equation}
where $\eps^{\mu}$ is the associated polarization vector to the current $J_{\mu}(x)$. Here, the hadronic tensor $W_{\mu\nu}$ is defined as the spin-averaged, squared modulus of the matrix element in Eq.~\eqref{eq:hadronicamp} combined with the total hadronic contribution to the phase-space element of Eq.~\eqref{eq:diffxsec}. Namely, 
\begin{align}
\label{eq:hadronictensor}
W_{\mu\nu} &\equiv \overbar{\sum_{\text{spins}}}\sum_{X}\int d\Pi_{X}(2\pi)^{4}\delta^{4}\left(p + k - k' - p_{X}\right)\bra{P}J_{\mu}(0)\ket{X}_{c}\bra{X}J_{\nu}(0)\ket{P}_{c} \nonumber\\
& = \overbar{\sum_{\text{spins}}}\int d^{4}x e^{iq\cdot x}\bra{P}\left[J_{\mu}(x),J_{\nu}(0)\right]\ket{P}_{c}.
\end{align}
This result is arrived at by considering the completeness of hadronic states, translation invariance, and the timelike positivity of $q^{0}$. The subscript $c$ denotes connected matrix elements. For simplicity, we have suppressed the additional quantum numbers labeling the states since they are unimportant for our discussion. The hadronic tensor can be related to the Compton amplitude $T_{\mu\nu}$ through the optical theorem in the special case of forward scattering:
\begin{equation}
\label{eq:hadronicCompton}
2\text{Im}\left[T_{\mu\nu}\right] = W_{\mu\nu},
\end{equation}
where the Compton amplitude $T_{\mu\nu}$ is defined as 
\begin{equation}
\label{eq:forwardCompton}
T_{\mu\nu} = i\overbar{\sum_{\text{spins}}}\int d^{4}x e^{iq\cdot x}\bra{P}T J_{\mu}(x) J_{\nu}(0) \ket{P}_{c}.
\end{equation}
The advantage in working with $T_{\mu\nu}$ is that it is more straightforward to calculate by virtue of the time-ordered product of currents, in addition to needing only the imaginary part. Given all of this, the structure of the hadronic current in terms of its constituents is still unknown. A way to circumvent this issue is to focus on the case of large momentum transfer $-q^{2} \equiv Q^{2} \gg M^{2}$, with $M$ being the proton mass. In fact, in this limit the matrix element in Eq.~(\ref{eq:forwardCompton}) can be calculated in a twist expansion using an operator product expansion or, equivalently, the parton model (see Ref.~\cite{LVdis} for an in-depth discussion of this point). In the following we adopt the parton-model picture in which interactions among the partons within the proton can be neglected due to asymptotic freedom.
Under these assumptions, the exchanged boson interacts with an asymptotically free parton of flavor $f$ carrying a longitudinal momentum fraction $\xi$ of the proton's momentum $p$ that has decohered from its surrounding environment. This allows one to assume the condition of incoherent scattering. The forward Compton amplitude $T_{\mu\nu}$ in the parton model can thus be written as
\begin{equation}
\label{eq:forwardComptonparton}
T_{\mu\nu} \approx i\overbar{\sum_{\text{spins}}}\sum_{f}\int d\xi f_{f}(\xi)\int d^{4}x e^{iq\cdot x}\bra{q_{f}(\xi p)}T J_{\mu}(x) J_{\nu}(0)\ket{q_{f}(\xi p)}_{c},
\end{equation}
where $f_{f}(\xi)$ are the parton distribution functions (PDFs). We now proceed to the discussion of how Lorentz violation affects the calculation of Eq.~\eqref{eq:forwardComptonparton} and the other physical quantities that appear in the cross section, Eq.~\eqref{eq:diffxsec}. 

\subsection{Lorentz-Violating Effects}
Lorentz-violating effects in unpolarized $e$-$p$ DIS were first studied in Ref.~\cite{LVdis} in the context of HERA collider data \cite{hera}, and we refer the reader to these documents to complement the discussion that follows. As in Ref.~\cite{LVdis}, we use the mSME to describe the inclusion of Lorentz-violating tree-level effects that control the hard interaction in unpolarized $e$-$p$ DIS. For simplicity, we focus on effects emanating in the high-energy regime of the mSME with the restriction of electron and $u, d$ quark flavor content, the latter consideration owing to the dominant flavor content of the proton. Note that, in the hard interaction, the vector boson is exchanged in a $t$-channel diagram, implying a suppression of the $Z$ boson contribution; this situation is radically different in, e.g., the Drell-Yan processes in which the vector boson is exchanged in the $s$-channel ($q^2>0$). In light of this, we neglect altogether $Z$ boson effects. The dominant Lorentz-violating terms we consider are then
\begin{equation}
\label{eq:SMEextra}
\mathcal{L}_{\text{SME}} \supset 
  \tfrac{1}{2}i c_Q^{\mu\nu}\bar{Q}\gamma_{\mu}\overset{\text{\tiny$\leftrightarrow$}}D_{\nu}Q
+ \tfrac{1}{2}i c_U^{\mu\nu}\bar{U}\gamma_{\mu}\overset{\text{\tiny$\leftrightarrow$}}D_{\nu}U
+ \tfrac{1}{2}i c_D^{\mu\nu}\bar{D}\gamma_{\mu}\overset{\text{\tiny$\leftrightarrow$}}D_{\nu}D
-\tfrac{1}{4}\kappa_{F}^{\kappa\lambda\mu\nu}F_{\kappa\lambda}F_{\mu\nu} -\tfrac{1}{4}\kappa_{G}^{\kappa\lambda\mu\nu}G^{a}_{\kappa\lambda}G^{a}_{\mu\nu}  \;,
\end{equation}
where $\overset{\text{\tiny$\leftrightarrow$}} D_\nu = \overset{\text{\tiny$\leftrightarrow$}} \partial_\nu + 2 i \hat Q A_\nu$, $\hat Q$ is the charge operator, $Q$ denotes the left-handed $SU(2)$ quark doublet, $U$ and $D$ are the $SU(2)$ singlets, $\kappa_{F,G}^{\kappa\lambda\mu\nu}$ and $c_f^{\mu\nu}$ ($f=Q, U,D $) are the photon/gluon and quark coefficients for Lorentz violation, respectively. We work in a scenario in which the coefficients for Lorentz violation are generated far above the electroweak symmetry breaking scale, implying that all the mSME terms we consider have to be expressed in terms of $SU(2)\times U(1)$ multiplets. 

It is important to mention that not all coefficients that appear in the SME are observable because  the theory is invariant in form  under spacetime-dependent field redefinitions and a change of coordinates~\cite{SME1, SME2, fermionobservables, mattergravity}.
As an example, the fermion field redefinition of the form $\psi(x) \to \exp\left[i f(x)\right]\psi(x)$ with $f(x) = a_{\mu}x^{\mu}$ can be used to remove completely the term $-a_\mu\bar \psi \gamma^\mu \psi$ which appears in ${\cal L}_{\text{QED}}^{\text{ext.}}$ in Eq.~\eqref{eq:LVQED}. However, in presence of more than one fermion species, it is in general not possible to remove all $a_f^\mu$ terms and some combination of them remains observable. Of more relevance to us, fermion field redefinitions of the form $\psi(x) \to [1 + v(x) \cdot \Gamma] \psi(x)$ with $\Gamma = \{\gamma^\alpha,\; \gamma^5 \gamma^\alpha,\; \sigma^{\alpha\beta}\}$ can be used to eliminate the antisymmetric part of all $c_\psi^{\mu\nu}$ terms that appear in Eq.~\eqref{eq:SMEextra}; moreover, it should be pointed out that these coefficients can be taken to be traceless because terms proportional to the Minkowski metric $\eta^{\mu\nu}$ do not violate Lorentz symmetry. 

The question of coordinate choice is more subtle. It is straightforward to show that the coordinate transformation $x^\mu \to x^\mu -\frac{1}{2} \kappa^{\alpha\mu}{}_{\alpha\nu} x^\nu$ (where $\kappa$ is a generic constant tensor) implies
\begin{align}
(\kappa_{F,G})^{\alpha\mu}{}_{\alpha\nu} &\to (\kappa_{F,G})^{\alpha\mu}{}_{\alpha\nu}-\kappa^{\alpha\mu}{}_{\alpha\nu} \; ,\\
c^{\mu\nu}_f &\to c^{\mu\nu}_f +\tfrac{1}{2} \kappa^{\alpha\mu\hphantom{\alpha}\nu}_{\hphantom{\alpha\mu}\alpha} \; . 
\end{align}
Taking into account that the traces $(\kappa_{F,G})^{\alpha\mu}{}_{\alpha\nu}$ contain all the independent components of the tensors $\kappa_{F,G}$, it is clear that this change of coordinates can be used to remove completely one amongst the coefficients $\kappa_{F}$, $\kappa_G$ and $c_f$ (where $f$ includes {\em all} fermion species). 
In our analysis, we choose coordinates in which the photon coefficient $\kappa_F$ vanishes. With this choice of coordinates the electron and proton coefficients\footnote{The proton $c^{\mu\nu}$ coefficients appear in the low-energy effective Chiral Lagrangian in which the proton is represented by a fundamental field (see also the discussion in Ref.~\cite{hadronic1}).} are experimentally tightly constrained~\cite{datatables} and have negligible impact on $e$-$p$ DIS. We then focus on effects due to the $c_f$ coefficients for $u$ and $d$ quarks while ignoring, at the present time, the effect of Lorentz violation in the gluon sector ($\kappa_G$).

\begin{figure}[t]
\centering
\includegraphics[width=0.57 \linewidth]{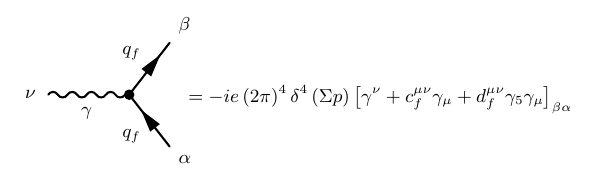}
\caption{Quark ($q_{f}$) interaction vertices from the model Lagrange density, Eq.~\eqref{eq:model}. The dot at the vertices denotes a modified vertex rule in the presence of Lorentz violation.}  \label{fig:feynrules} 
\end{figure}

\begin{figure}[t]
\centering
\includegraphics[width=0.4 \linewidth]{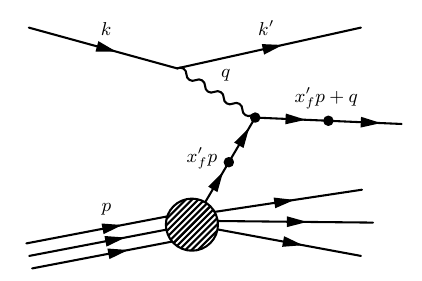}
\caption{Parton-model DIS with Lorentz-violating effects. A parton carrying a momentum fraction $x_f' = x - x_f$  propagates in a Lorentz-violating medium and interacts with an electron (see Ref.~\cite{LVdis}). These features affect the azimuthal scattering angle $\phi$ of the final-state electron. Dots indicate modified vertices and propagator insertions.\label{fig:DISparton}}
\end{figure}

Below the electroweak scale it is customary to express the fields $Q$, $U$ and $D$ in terms of the usual Dirac fields $\psi_u$ and $\psi_d$. The model Lagrange density in the massless limit is then given by
\begin{align}
\label{eq:model}
\mathcal{L} = \sum_{f=u,d}\tfrac{1}{2}\bar{\psi}_{f} (\gamma^{\nu} + c_f^{\mu\nu} \gamma_\mu + d_f^{\mu\nu} \gamma_5 \gamma_\mu ) i\overset{\text{\tiny$\leftrightarrow$}}{D}_{\nu}\psi_{f} \; ,
\end{align}
with 
\begin{align}
\label{eq:quarkcoeffs}
c_u^{\mu\nu} &= (c_Q^{\mu\nu} + c_U^{\mu\nu})/2, \quad c_d^{\mu\nu} = (c_Q^{\mu\nu} + c_D^{\mu\nu})/2  \;, \nonumber\\
d_u^{\mu\nu} &= (c_Q^{\mu\nu} - c_U^{\mu\nu})/2, \quad d_d^{\mu\nu} = (c_Q^{\mu\nu} - c_D^{\mu\nu})/2  \;.
\end{align}
As discussed above, these coefficients control the magnitude of Lorentz violation and can be taken as traceless and symmetric~\cite{mattergravity,fermionobservables}. For simplicity, we also assume they are constants in a given inertial observer frame, which ensures energy-momentum conservation through invariance under time translations.  At leading-order in the coefficients, the fermion propagator takes the form\footnote{The calculation of the inverse is most easily performed using the fact that $(\slashed{p}+\gamma_5 d_f^{\gamma p})(\slashed{p}-\gamma_5 d_f^{\gamma p})(\slashed{p}-\gamma_5 d_f^{\gamma p})(\slashed{p}+\gamma_5 d_f^{\gamma p}) = (p^2 + 2 \gamma_5 d_f^{pp})(p^2 - 2 \gamma_5 d_f^{pp}) = p^4$ up to higher orders in $d_f^{\mu\nu}$. This leads to the second term in the round bracket.} 
\begin{align}
\raisebox{-0.26cm}{\includegraphics[width=0.15 \linewidth]{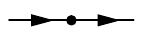}}
\quad&= 
\frac{i}{(\eta^{\mu\nu}+ c_f^{\mu\nu} + \gamma_5 d_f^{\mu\nu}) \gamma_\mu p_\nu } \nonumber\\
& = i \frac{\slashed{p} + c_f^{\mu\nu} \gamma_\mu p_\nu}{p^2+2 c_f^{\mu\nu} p_\mu p_\nu}
+ i \gamma_5 \frac{1}{p^2} \left( d_f^{\mu\nu} \gamma_\mu p_\nu - 2i p^\mu  \sigma_{\mu\nu} d_f^{\nu \alpha} p_\alpha  \frac{\slashed{p}}{p^2} \right) \; ,
\label{eq:quarkprop}
\end{align}
\noindent where $\sigma_{\mu\nu} = \tfrac{i}{2}[\gamma_\mu,\gamma_\nu]$ and the associated vertex rule is shown in Fig.~\ref{fig:feynrules}. As in the Lorentz-invariant case, each vertex contributes a momentum-conserving delta function $\delta^{4}\left(\Sigma p\right)$ by virtue of preserved translational invariance in the mSME. In general, it is not possible to consider scenarios in which just one of the four coefficient combinations in Eqs.~\eqref{eq:quarkcoeffs} is present. However, in the particular case of \ti{unpolarized} scattering mediated by a virtual photon, the coefficients $d_{f}^{\mu\nu}$ do not produce observable effects. This is because the leptonic contribution to Eq.~\eqref{eq:diffxsec} yields the usual symmetric lepton tensor $L^{\mu\nu} = 2\left(k^{\mu}k'^{\nu} + k^{\nu}k'^{\mu} - (k\cdot k')\eta^{\mu\nu}\right)$. Due to the presence of an explicit $\gamma_5$ in Eq.~\eqref{eq:quarkprop}, the terms proportional to $d_f^{\mu\nu}$ produce an antisymmetric tensor which vanishes once contracted $L^{\mu\nu}$. Note that the coefficients $d_f^{\mu\nu}$ can produce observable effects if the DIS process is mediated by a $Z$ boson or if polarized observables are of interest.  For our purposes, the only physically relevant coefficients are $c_{f}^{\mu\nu}$ and
they can be considered independently from each other. The unpolarized DIS process in the presence of Lorentz violation within the parton-model assumptions is thus depicted in Fig.~\ref{fig:DISparton}.\par

What has yet to be considered of the differential cross section Eq.~\eqref{eq:diffxsec} is the flux factor $F$ and final-state electron phase-space element. Extra attention must be taken in the calculation of $F$ since in the presence of Lorentz violation $F$ is frame dependent and particle trajectories are affected due to a modified energy-momentum dispersion relation and group velocity \cite{xsections}. Since we are working in the DIS limit and employing the parton model, the flux is calculated between the electron and parton. Because we are only considering modifications to freely-propagating partons in the initial state, it is reasonable to neglect the group-velocity modifications of $F$ \cite{LVdis,xsections}. The flux then takes the conventional form
\begin{align}
\label{eq:flux}
F &\simeq 4k\cdot \xi p \simeq 2\xi s ,
\end{align}
where $s\simeq 2k\cdot p$. Note that the factor $\xi$ that appears must be included in the integrand of Eq.~\eqref{eq:forwardComptonparton}. The last remaining piece of Eq.~\eqref{eq:diffxsec} is the phase-space element for the final-state electron. A change of variables reveals
\begin{equation}
\label{eq:electronphasespace}
\int \fr{d^{3}k',(2\pi)^{3}}\fr{1,2E_{k'}}  = \int \fr{d\phi,2\pi}\int dx dy\fr{ys,(4\pi)^{2}},
\end{equation}
where we have introduced the canonical Bjorken variables (neglecting electron and quark masses) $x = -q^{2}/(2p\cdot q) \simeq  k\cdot k'/p\cdot q$ and $y = p\cdot q/p\cdot k$. As we mentioned in the caption of Fig.~\ref{fig:DISparton}, the dependence on the final-state electron scattering angle $\phi$ is now nontrivial, which is why we leave the angular integral in Eq.~\eqref{eq:electronphasespace}. 
The triply differential cross section can be shown to take the following form:
\begin{align}
\label{eq:photonex}
\frac{d^{3}\sigma}{dxdyd\phi} &= \frac{\alpha^{2}s}{q^{4}}\sum_{f=u,d,s,\cdots}q_{f}^{2}xf_{f}(x)\left[1+(1-y)^{2}\right] \nonumber\\ 
& +  \frac{\alpha^{2}}{q^{4}}\sum_{f=u,d}q_{f}^{2}xf_{f}(x)\left[\left[C''\right] - \frac{2(1+(1-y)^{2})}{y}\left(\left[C'\right] + \left(\frac{1}{x} + \frac{d\ln f_{f}(x)}{dx}\right)[C]\right)\right],
\end{align}
where
\begin{align}
\label{eq:coeffs}
& [C] = c_{f}^{\mu\nu}\left[q_{\mu}q_{\nu} + x\left(q_{\mu}p_{\nu} + q_{\nu}p_{\mu}\right) + x^{2}p_{\mu}p_{\nu}\right],\nonumber\\
& [C'] = c_{f}^{\mu\nu}\left(p_{\mu}q_{\nu} + p_{\nu}q_{\mu} + 2xp_{\mu}q_{\nu}\right),\nonumber\\
& [C''] = \frac{2y}{x}[C] + c_{f}^{\mu\nu}\left[4\left(k'_{\mu}p_{\nu} + p_{\mu}k'_{\nu}\right) + \frac{4}{x}(1-y)k_{\mu}k_{\nu} + 4\left(1-y\right)\left(k_{\mu}p_{\nu} + p_{\mu}k_{\nu}\right) - 4xyp_{\mu}p_{\nu} - \frac{4}{x}k'_{\mu}k'_{\nu}\right].
\end{align}
The first line of Eq.~\eqref{eq:photonex} is the leading-order Lorentz-invariant contribution. Note that the sum for this portion includes all parton species. The second line, which is proportional to $c_{f}^{\mu\nu}$, is the contribution from the dominant Lorentz-violating effects on the $u$ and $d$ quarks. The appearance of nontrivial dependence on the final-state electron azimuthal angle $\phi$ as well as scaling violations is now transparent. In addition, the cross section now depends separately on the electron and proton energies and not just on the CM energy $s=4 E_e E_p$. This can be made more clear by defining the kinematics in the laboratory frame with the choice of electron propagation direction defining the laboratory $z$ axis. The four momentum of the incident electron, proton, and scattered electron then take the form $k^{\mu} =  E_{e}\left(1,+\hat{z}\right)$, $p^{\mu} = E_{p}\left(1,-\hat{z}\right)$, and $k'^{\mu} = E'_{e}(1,\hat{k}')$ with $\hat{k}' = \left(\sin\theta\cos\phi,\sin\theta\sin\phi,\cos\theta\right)$, respectively. 

\subsection{Sidereal Time Dependence and the Sun-Centered Celestial-Equatorial Frame}
We are now in the position to pick a frame in order to estimate attainable bounds on the coefficients $c_{f}^{\mu\nu}$. In choosing a frame to analyze an expression like Eq.~\eqref{eq:photonex}, special care must be taken because the coefficients for Lorentz violation depend on the choice of observer frame. This implies that an Earth-based experiment will exhibit a sidereal time dependence in the cross section. It is therefore important to work initially in a suitable (approximately) inertial frame. The standard choice of frame used for reporting bounds on the coefficients for Lorentz violation is known as the Sun-centered celestial-equatorial frame (SCF) \cite{sccef,sccef2,datatables}. This frame is effectively inertial over the time scale of most Earth-based experiments. We remind the reader that we take the coefficients to be constant in this frame. The transformation from the SCF frame to the laboratory frame is reasonably approximated and achieved by a series of rotations \cite{sccef,sccef2}. The net rotation is given by the product of the following matrices \cite{LVdis,xsections}:
\begin{align}
\label{eq:rotation}
\mathcal{R} = \begin{pmatrix}\pm 1 & 0 & 0 \\ 0 & 0 & 1 \\ 0 & \mp 1 & 0\end{pmatrix} \begin{pmatrix}\cos\psi & \sin\psi & 0 \\ -\sin\psi & \cos\psi & 0 \\ 0 & 0 & 1\end{pmatrix} \begin{pmatrix}\cos\chi\cos\omega_{\oplus} T_{\oplus} & \cos\chi\sin\omega_{\oplus} T_{\oplus} & -\sin\chi \\ -\sin\omega_{\oplus} T_{\oplus} & \cos\omega_{\oplus} T_{\oplus} & 0 \\ \sin\chi\cos\omega_{\oplus} T_{\oplus} & \sin\chi\sin\omega_{\oplus} T_{\oplus} & \cos\chi\end{pmatrix}.
\end{align}
In this expression, $\omega_{\oplus} \simeq 2\pi$/(23 \text{hr} 56 \text{min}) is the Earth's sidereal frequency which accompanies the local sidereal time $T_{\oplus}$ \cite{localsidereal}. The angle $\chi$ and $\psi$ refer to the colatitude of the laboratory and the orientation of the electron beam momentum $\hat{k}$ relative to the East cardinal direction, respectively. The last rotation involving only unit and null entries is an inversion of coordinates to orient the Earth-frame $z$ direction, which is initially perpendicular to the surface of the Earth prior to this rotation, along the direction of the electron beam momentum. The consideration of two potential orientations as indicated by the $\pm$ and $\mp$ will be useful for the analysis of EIC simulated data in Sec.~\ref{sec:EIC}. Under the rotation of Eq.~\eqref{eq:rotation}, the coefficients $c_{f}^{\mu\nu}$ are transformed via the standard transformation rules:
\begin{equation}
\label{eq:rotatedcoeffs}
c_{f,\text{lab.}}^{\mu\nu} = \begin{cases} \mathcal{R}^{ik}\mathcal{R}^{jl}{c_{f}}_{kl},\quad \mu,\nu = i,j \in \{1,2,3\} \\ 
\\
\mathcal{R}^{ik} {c_{f}}_k{}^0,\quad \mu,\nu = i,0 ,\end{cases}
\end{equation}
where the sum over repeated indices is implied. In view of this transformation, it is clear that once the coefficients have been transformed to the laboratory frame, time dependence in the differential cross section becomes manifest. Direct observation reveals that sinusoidal oscillations involving the first and second harmonics of $\omega_{\oplus}$ accommodate some of the coefficients under this transformation. This implies the differential cross section given in Eq.~\eqref{eq:photonex} oscillates around the (shifted) SM result at the first and second harmonics of $\omega_{\oplus}$ and therefore represents a distinct signal for Lorentz violation. Considering the properties of $c_{f}^{\mu\nu}$ and further inspection of Eq.~\eqref{eq:rotatedcoeffs} reveals that, for a given flavor $f$, only six combinations of the nine independent components of $c_{f}^{\mu\nu}$ contribute to the inherited time dependence of the cross section. More specifically, the coefficients $c_{f}^{TX}$, $c_{f}^{TY}$, $c_{f}^{XZ}$, and $c_{f}^{YZ}$ involve first harmonics of $\omega_{\oplus}$, while $c_{f}^{XY}$ and the combination $c_{f}^{XX} - c_{f}^{YY}$ involve second harmonics of $\omega_{\oplus}$ where $\{T,X,Y,Z\} = \{0,1,2,3\}$. The other three combinations, $c_{f}^{TZ}$, $c_{f}^{ZZ}$, and $c_{f}^{TT} \equiv c_{f}^{XX} + c_{f}^{YY} + c_{f}^{ZZ}$ contribute only as constant offsets to the SM result. As such, these coefficients will not be relevant to our primary analysis, but we will provide rough estimates of their influence in Section~\ref{sec:timeind}. The first estimates of bounds on the time-dependent coefficients attainable from existing HERA data were discussed in Ref.~\cite{LVdis}. In the next section we perform a similar analysis, but using simulated data to provide predictions for the sensitivities to Lorentz violation at the future EIC.

\section{Estimated constraints for the EIC}\label{sec:EIC}
\subsection{Collider Details and Bound Extraction Procedure}
The expected sensitivities to the quark-sector coefficients for Lorentz violation $c_{f}^{\mu\nu}$ can be calculated by means of simulated data describing the kinematical coverage and experimental uncertainties of the proposed EIC designs. 
To start the discussion in this section, we briefly discuss the relevant features of the JLEIC and eRHIC with regards to our numerical analysis. The current baseline JLEIC design concept features a new figure-eight collider ring with two interaction points and a beam luminosity on the order of   $10^{34}~\text{cm}^{-2}~\text{s}^{-1}$ \cite{EICsummary}. The projected beam energy range for the electrons and protons is $ 3 \leq E_{e} \leq 12$ GeV and $20 \leq E_{p} \leq 100$ GeV, respectively, leading to a CM energy range of roughly $15 \leq\sqrt{s}\leq 70$ GeV. The collider colatitude is $\chi \approx \ang{52.9}$ with electron beam orientations $\psi \approx \ang{47.6}, \ang{-35.0}$ North East (NE) that we henceforth refer to as JLEIC1 and JLEIC2, respectively. The baseline eRHIC concept, in contrast, utilizes the existing relativistic heavy-ion collider ring with interaction points at the STAR and PHENIX detectors and a luminosity on the order of $10^{33}~\text{cm}^{-2}~\text{s}^{-1}$ \cite{eRHICdesign}. The projected beam energy range for the electrons and protons is $ 5 \leq E_{e} \leq 20$ GeV and $50 \leq E_{p} \leq 250$ GeV, respectively, leading to a CM energy range of roughly $30 \leq\sqrt{s}\leq 140$ GeV. The colatitude of the collider is $\chi \approx \ang{49.1}$ and electron beam orientations are $\psi \approx \ang{-78.5}, \ang{-16.8}$ NE for the PHENIX and STAR detectors, respectively, and we henceforth refer to these configurations as eRHIC1 and eRHIC2, respectively.\par

As suggested in Sec.~\ref{sec:phenom}, our primary interest is to explore the sensitivities of the 12 coefficients that induce sidereal time dependence in the differential cross section given by Eq.~\eqref{eq:photonex}. The focus on this subset of coefficients is mainly due to two reasons. On one hand, the fits which yield the PDFs are performed using several time-averaged cross sections for several processes, including $e$-$p$ DIS: any Lorentz-violating effects on the time-averaged cross section would be (at least partially) reabsorbed into the fitted PDFs. It is, therefore, difficult to extract reliable bounds on coefficients which contribute to time-averaged quantities. On the other hand, the extraction of bounds from time-dependent effects is favored because the SM-induced sidereal time variation is exactly null and 100\% systematic uncertainties do not affect, to a very good extent, the resulting bounds. The reason for this latter fact is that 100\% correlated errors will move all sidereal time bins in the same direction without inducing any spurious time dependence (which is obviously generated by statistical uncertainties). In particular, both systematic theory and experimental uncertainties are almost 100\% correlated; this is especially relevant for the former because they can be sizable. We should mention that it will be necessary to monitor carefully the time dependence of the luminosity measurement in order to obtain the sidereal bins integrated luminosities with correlated uncertainties. In general, systematic uncertainties emanating from day/night differences are distinguishable from sidereal effects because the former undergo an shift of approximately 4 min/day. If the data are taken over a longer time period (i.e., of order one year), as is typical for these kinds of experiments, the day/night systematics can be well suppressed compared to the sidereal counterpart. Otherwise, it may be possible to separate and distinguish these effects in the data.

The method of analysis used to place bounds is similar to what was done in Ref.~\cite{LVdis} (to which we refer the reader for further details), but has a few key differences. The starting point in Ref.~\cite{LVdis} were the combined Zeus/H1 DIS results presented at fixed $x$ and $Q^2$. Note that experiments measure cross sections in bins of $x$ and $Q^2$ and ascribe the average to the center of the two-dimensional bin. For each $x$ and $Q^2$ value, we generated 1000 Gaussian-distributed pseudoexperiments, each of which describes the potential outcome of splitting the HERA dataset into four bins in sidereal time.  We also required the weighted average of the binned cross sections to be identical to the measured one: in this way, each pseudoexperiment simulates what the actual splitting of the (already observed) events in sidereal bins might look like. The expected upper limit on the coefficients for Lorentz violation of interest is then given by the median of the upper limits calculated for each pseudoexperiment. We extract the latter by building the following chi-square:
\begin{align}
\chi^2_i (x,Q^2,c_f^{\mu\nu}) = \sum_{n,m=1}^{n_{\rm bins}} \Big[\sigma_{\rm th} (x,Q^2,c_f^{\mu\nu},n) - \sigma_{\rm exp}^i (x,Q^2,n) \Big] \;  C_{nm}^{-1} \; \Big[ \sigma_{\rm th} (x,Q^2,c_f^{\mu\nu},m) - \sigma_{\rm exp}^i (x,Q^2,m)\Big],
\end{align}
where $n_{\rm bins}$ is the number of sidereal time bins, $\sigma_{\rm th} (x,Q^2,n)$ is the theoretical cross section integrated over the $n$th sidereal time bin,  $\sigma_{\rm exp}^i (x,Q^2,n)$ is the corresponding experimental cross section as it appears in the $i$th pseudoexperiment, and $C_{nm}$ is the $n_{\rm bins}\times n_{\rm bins}$ covariance matrix. The latter is the sum of two contributions: the statistical part is diagonal and is rescaled by a factor equal to the square root of the number of bins with respect to the statistical error of the original measurement, and we consider the two extreme cases of 0\% and 100\% correlation for the systematic contributions. In contrast, for this analysis there are no existing data for the EIC and we simply sample the SM cross section in each sidereal time bin. We use the software ManeParse~\cite{Maneparse1, Maneparse2}, and the CT10 set~\cite{CT10} in particular, for the quark PDFs. Each coefficient is bounded independently by setting the others to zero, which is in accordance with accepted procedure \cite{datatables}. In addition to the extraction of bounds based on a chi-square for each individual measurement, we also construct a global chi-square over the entire dataset for each collider and the two respective detector locations and orientations:
\begin{align}
\chi^2_i (c_f^{\mu\nu}) = \sum_{x,Q^2} \chi^2_i (x,Q^2,c_f^{\mu\nu}) \; .
\end{align}

\begin{table}[t]
\begin{tabular}{|lccccccc|}
\hline
& {\bf JLEIC} 
&\;\;& {\bf JLEIC} 
&\;\;&  {\bf eRHIC} 
&\;\;&  {\bf eRHIC} 
 \\
&  {\bf Individual} 
&\;\;& {\bf Global} 
&\;\;&  {\bf Individual} 
&\;\;&  {\bf Global} 
 \\[0.1cm] \hline \\[-0.3cm]
$|c_{u}^{TX}|$ & 0.04\;[0.2] && 0.03\;[0.1] && 0.1\;[0.5]  && 0.04\;[0.3]\\ 
               & 0.04\;[0.2] && 0.02\;[0.1] && 0.09\;[0.3] && 0.03\;[0.2]\\ 
$|c_{u}^{TY}|$ & 0.04\;[0.2] && 0.03\;[0.1] && 0.1\;[0.5]  && 0.04\;[0.3]\\ 
               & 0.04\;[0.2] && 0.02\;[0.1] && 0.09\;[0.3] && 0.03\;[0.2]\\ 
$|c_{u}^{XZ}|$ & 0.07\;[0.4] && 0.05\;[0.2] && 0.2\;[0.7]  && 0.05\;[0.4]\\
               & 0.08\;[0.4] && 0.05\;[0.3] && 0.4\;[2.0]   && 0.1\;[0.8]\\ 
$|c_{u}^{YZ}|$ & 0.07\;[0.4] && 0.05\;[0.2] && 0.2\;[0.7]  && 0.05\;[0.4]\\ 
               & 0.09\;[0.4] && 0.05\;[0.3] && 0.4\;[2]   && 0.1\;[0.8]\\ 
$|c_{u}^{XY}|$ & 0.3\;[1] && 0.2\;[0.9]   && 0.5\;[2]   && 0.1\;[1]\\ 
               & 0.1\;[0.7] && 0.08\;[0.4] && 0.2\;[0.7]  && 0.06\;[0.4]\\ 
$|c_{u}^{XX} - c_{u}^{YY}|$ & 0.2\;[1] && 0.1\;[0.7] && 1.0\;[5]   && 0.4\;[3] \\
                            & 0.2\;[1] && 0.2\;[0.8] && 0.9\;[4]  && 0.3\;[2]
\\[0.1cm] \hline \\[-0.3cm]
$|c_{d}^{TX}|$ & 0.7\;[4] && 0.3\;[2] && 1\;[10]  && 0.4\;[4]\\ 
               & 0.7\;[3] && 0.3\;[2] && 0.9\;[6] && 0.3\;[3]\\ 
$|c_{d}^{TY}|$ & 0.7\;[4] && 0.3\;[2] && 1\;[9] && 0.4\;[4]\\ 
               & 0.6\;[3] && 0.3\;[2] && 0.9\;[7] && 0.3\;[3]\\ 
$|c_{d}^{XZ}|$ & 1\;[6] && 0.6\;[4] && 2\;[10] && 0.5\; [5]\\
               & 1\;[7] && 0.7\;[5] && 4\;[30] && 1 \;[10]\\ 
$|c_{d}^{YZ}|$ & 1\;[6] && 0.6\;[4] && 2\;[10] && 0.5\;[5]\\ 
               & 1\;[8] && 0.7\;[5] && 4\;[30] && 1\;[10]\\ 
$|c_{d}^{XY}|$ & 5\;[20] && 2\;[10] && 5\;[30] && 1\;[10]\\ 
               & 2\;[10] && 1\;[7] && 2\;[10] &&  0.5\;[6]\\ 
$|c_{d}^{XX} - c_{d}^{YY}|$ & 4\;[20] && 2\;[10] && 10\;[100] && 4\;[40] \\
                            & 4\;[20] && 2\;[10] && 10\;[70] && 3\;[30]\\ 
\hline
\end{tabular}
\caption{Expected best individual and global bounds for the JLEIC and eRHIC. All bounds are given in units of $10^{-5}$. The bounds with brackets correspond to the case of uncorrelated systematic uncertainties between binned data, and the bounds without brackets correspond to assuming 100\% correlation between systematic uncertainties. For each coefficient magnitude, we give the bounds for both electron beam orientations (see the caption in Tables~\ref{tab:JLEICallbounds}-\ref{tab:eRHICallbounds} for further information). \label{tab:all}}
\end{table}

\subsection{Numerical Results: Individual and Global Bounds for Time-Dependent Coefficients}
Datasets of simulated reduced cross sections with associated uncertainties over a range of $(E_{e}, E_{p})$ values characteristic of the JLEIC and eRHIC are used to extract the individual and global bounds. The datasets where generated using HERWIG 6.4~\cite{Bahr:2008pv, Bellm:2015jjp} at NLO and estimates of detector systematics were based off of the HERA collider~\cite{hera}. The JLEIC dataset comprises a total of 726 measurements covering the range $x \in \left(9\times 10^{-3}, 9\times 10^{-1}\right), Q^{2} \in \left(2.5, 2.2\times 10^{3}\right)$~GeV$^{2}$ with electron beam energies $E_{e} = 5, 10$ GeV and proton beam energies $E_{p} = 20, 60, 80, 100$ GeV. The DIS cross sections have been evaluated at next-to-leading order including power corrections stemming from higher twist and target mass effects~\cite{Accardi:2016qay}. These data correspond to an integrated luminosity of $100$ fb$^{-1}$, which represents roughly one year of data taking for the JLEIC. These data come with an overall point-to-point systematic uncertainty of $0.5\%$ for Bjorken $x < 0.7$ and $1.5\%$ for $x>0.7$ as well as a $1\%$ luminosity error. The dataset for the eRHIC comprises 1488 measurements covering the range $x \in \left(1\times 10^{-4}, 8.2\times 10^{-1}\right), Q^{2} \in \left(1.3, 7.9\times 10^{3}\right)$~GeV$^{2}$ with electron beam energies $E_{e} = 5, 10, 15, 20$ GeV and proton beam energies $E_{p} = 50, 100, 250$ GeV. The integrated luminosity is $100$ fb$^{-1}$ and represents roughly 10 years of data taking when accounting for the eRHIC luminosity. These data come with an overall $1.6\%$ point-to-point systematic uncertainty and a $1.4\%$ luminosity error. This dataset allows an investigation into a wider kinematical range. Taking the JLEIC and eRHIC data as a whole, the CM energy range is in total approximately $28 \leq \sqrt{s} \leq 141$~GeV, which is fairly representative of the expected full CM energy range of the EIC as in Section~\ref{sec:intro}.

\begin{figure}[t]
\centering
\includegraphics[width=0.99 \linewidth]{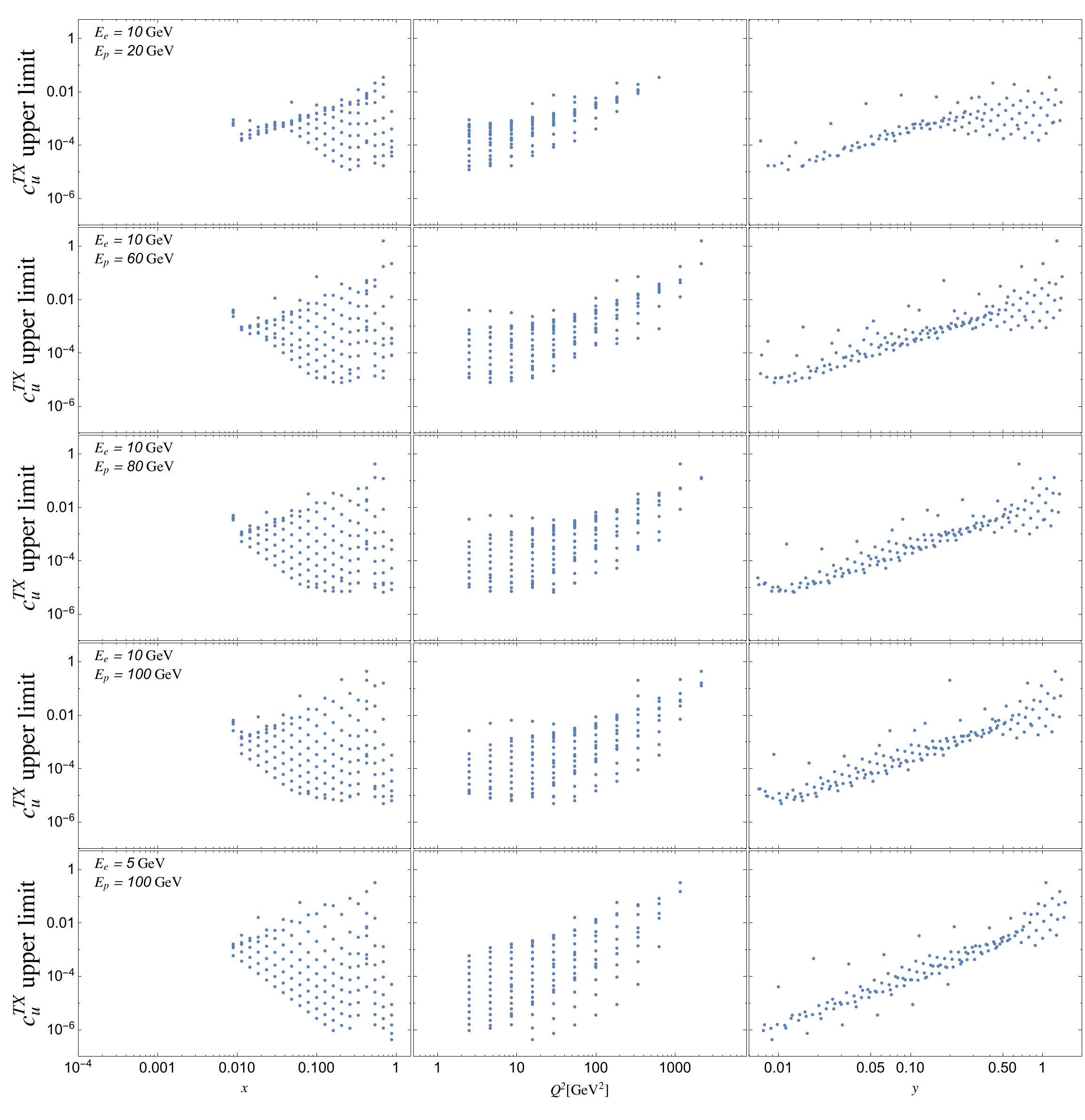}
\caption{Distribution of $|c_{u}^{TX}|$ correlated upper limits for JLEIC1. The median value for the collection of pseudoexperiments is plotted against the variables $x$, $Q^{2}$, $y$ for electron energies $E_{e} = 5, 10$ GeV and proton energies $E_{p} = 20, 60, 80, 100$ GeV.\label{fig:JLEIC1}}
\end{figure}

\begin{figure}[t]
\centering
\includegraphics[width=0.99\linewidth]{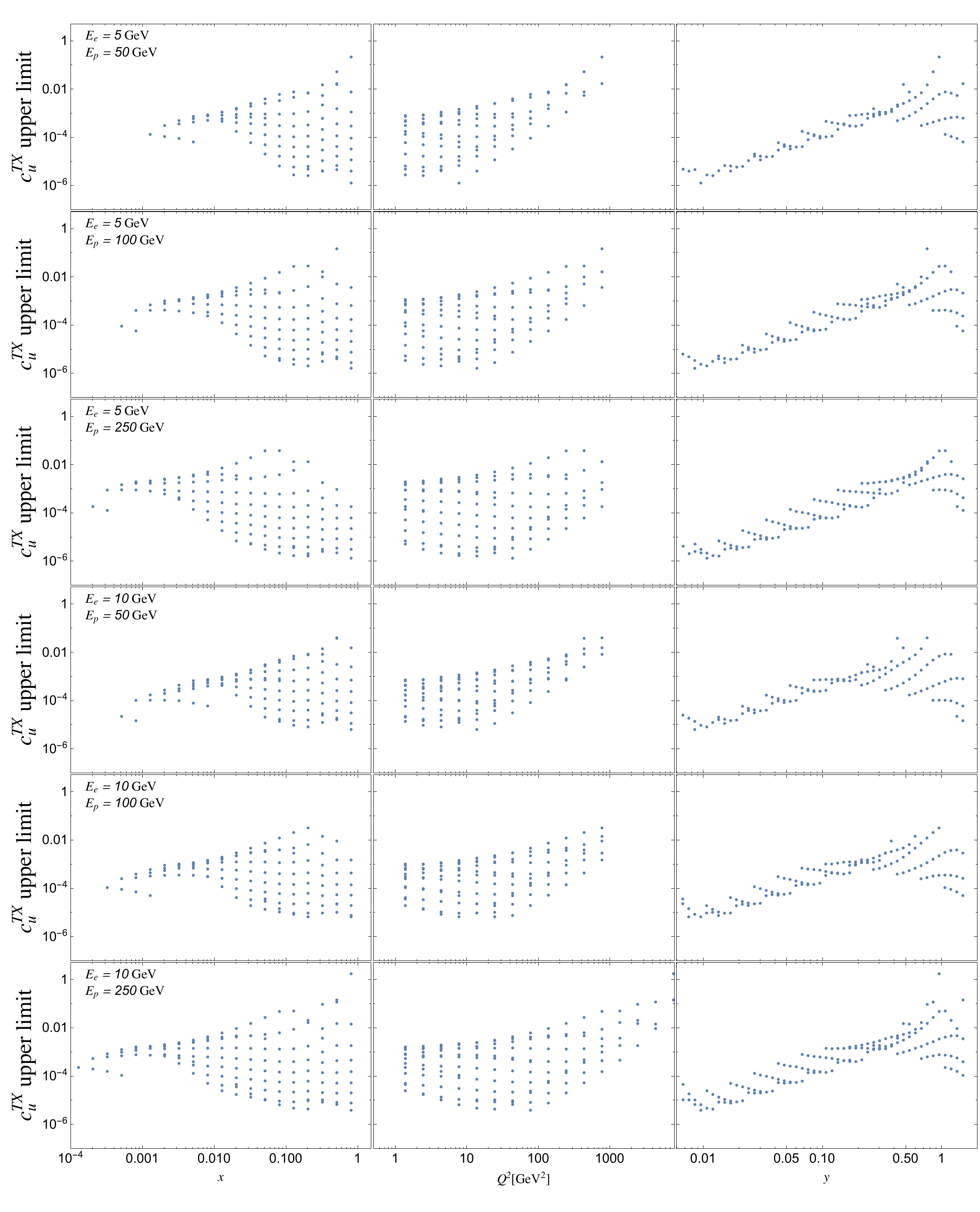}
\caption{Distribution of $|c_{u}^{TX}|$ correlated upper limits for eRHIC1. The median value for the collection of pseudoexperiments is plotted against the variables $x$, $Q^{2}$, $y$ for electron energies $E_{e} = 5, 10$ GeV and proton energies $E_{p} = 50, 100, 250$ GeV.\label{fig:eRHIC1}}
\end{figure}

\begin{figure}[t]
\centering
\includegraphics[width=0.99\linewidth]{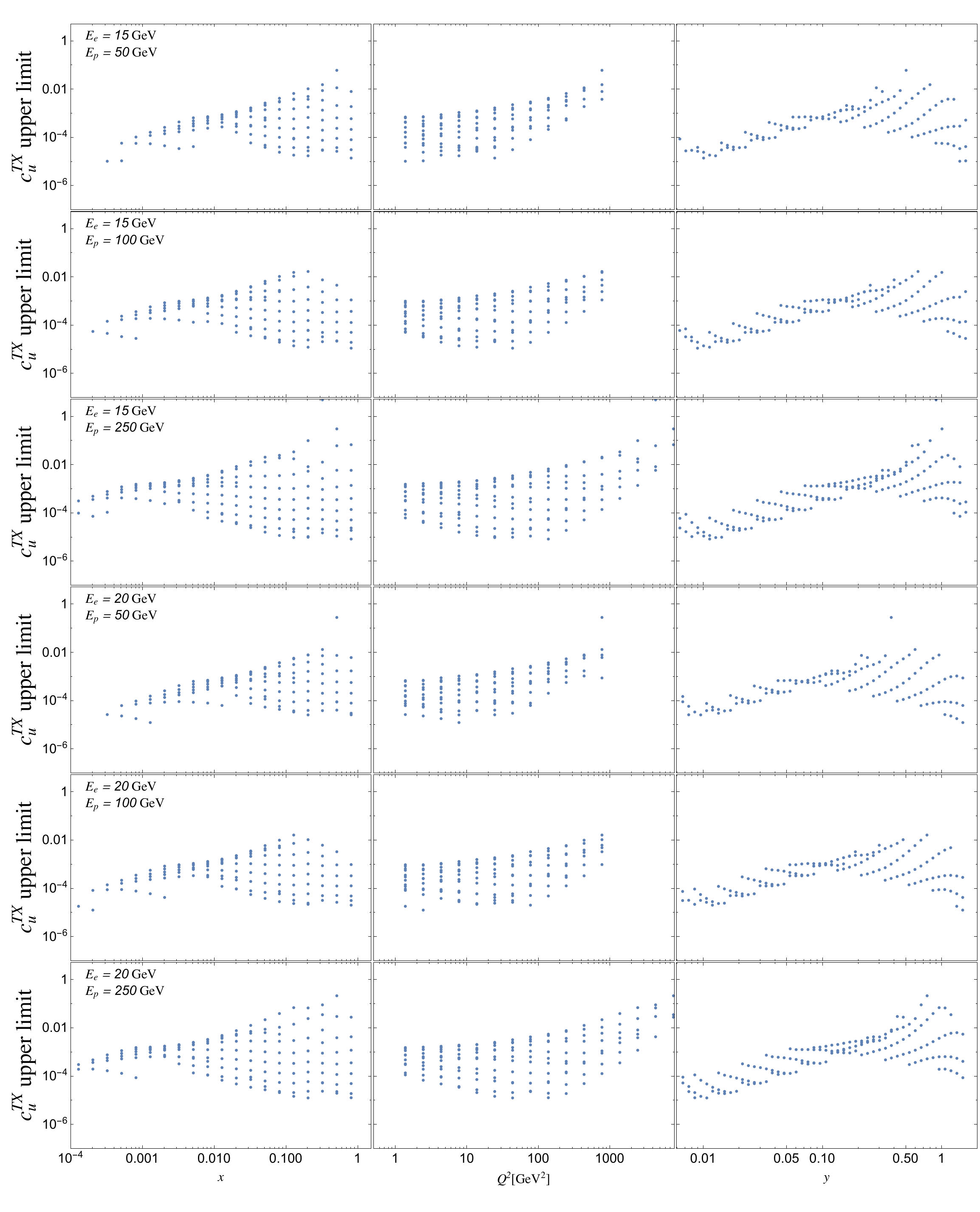}
\caption{Distribution of $|c_{u}^{TX}|$ correlated upper limits for eRHIC1. The median value for the collection of pseudoexperiments is plotted against the variables $x$, $Q^{2}$, $y$ for electron energies $E_{e} = 15, 20$ GeV and proton energies $E_{p} = 50, 100, 250$ GeV.\label{fig:eRHIC2}}
\end{figure}

\begin{figure}[t]
\centering
\includegraphics[width=0.99\linewidth]{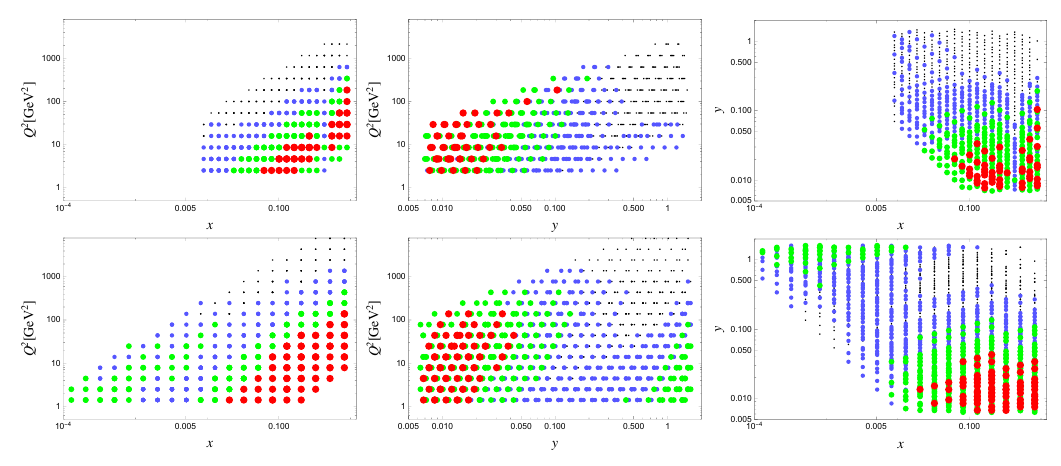}
\caption{Upper limits for $|c_{u}^{TX}|$ displaying ($x$, $Q^2$, $y$) correlations. Red, Green, Blue and Black dots corresponds to bounds below the $10^{-5}$, $10^{-4}$, $10^{-3}$ and $10^{-2}$ level, respectively. Upper and lower panels correspond to JLEIC1 and eRHIC1, respectively. All upper limits assume 100\% bin-to-bin correlation between experimental systematic uncertainties. \label{fig:corr}}
\end{figure}

Our main results are given in Table~\ref{tab:all} where we present a compilation of the best individual and global bounds for both JLEIC and eRHIC. A detailed breakdown of the bounds we obtain is presented in Tables~\ref{tab:JLEICallbounds}-\ref{tab:eRHICallbounds} and in Figs.~\ref{fig:JLEIC1}-\ref{fig:corr}. In Tables~\ref{tab:JLEICallbounds}-\ref{tab:eRHICallbounds}, we present the best individual limits for the 12 coefficients inducing sidereal time dependence for the entire JLEIC and eRHIC datasets. To illustrate how these bounds vary over each of the datasets, we show as an example the distribution of best individual limits for the $c_{u}^{TX}$ coefficient for JLEIC1 in Fig.~\ref{fig:JLEIC1}, and eRHIC1 in Figs.~\ref{fig:eRHIC1}-\ref{fig:eRHIC2}. The dependence of our sensitivity on $x$, $y$ and $Q^2$ is shown in Fig.~\ref{fig:corr} for both JLEIC and eRHIC.

We begin with a discussion of the JLEIC results. Table~\ref{tab:JLEICallbounds} shows a general trend of increasing sensitivity (i.e., smaller bounds) for both cases of uncorrelated and $100\%$ correlated systematic uncertainties with increasing $E_{p}$ and decreasing $E_e$, as well as most sensitivity to the $u$ quark coefficients containing $(0,3) = (T, Z)$ indices. The latter fact can be inferred as a direct consequence of the rotation properties in Eq.~\eqref{eq:rotatedcoeffs} and by considering the difference in the $u$, $d$ quark charges. The former feature of an increasing sensitivity for larger values of $E_p$ at fixed $E_e$ implies more sensitivity at larger $s$, which is expected from considerations of the form of Eqs.~\eqref{eq:photonex} and \eqref{eq:coeffs}. In principle, we expect increased sensitivity at large $E_e$ as well, but this effect is shadowed by the experimental acceptance cut which requires a lower bound $y > y_c$ (with $y_c\sim 10^{-2}$) and the corresponding bound $Q^2 = s x y >  s x y_c = Q^2_c$. This lower bound implies that, at fixed $x$, lower $E_e$ allow points with smaller $Q^2$ which, all else being equal, yield much larger cross sections (remember that the electromagnetic contribution scales as $Q^{-4}$), smaller statistical uncertainties and an increased sensitivity to time-dependent Lorentz-violating effects. Interestingly, there is an even split between JLEIC1/JLEIC2 in terms of which configuration has better sensitivities, although the differences are small overall which is to be expected. Since all coefficients show a similar trend of increasing sensitivity at larger $E_p$, inspecting Fig.~\ref{fig:JLEIC1} for the $c_{u}^{TX}$ coefficient for a given orientation (JLEIC1 shown) is sufficient to determine the overall kinematical regions of most sensitivity. Generally speaking, the bounds showing the most sensitivity appear to come from the large $x$, low $Q^{2}$, and low $y$ region of the phase space, which is also consistent with the features of Eqs.~\eqref{eq:photonex} and \eqref{eq:coeffs}. These observations are further supported by examining the top panels in Fig.~\ref{fig:corr}, which shows correlations between these variables for the special case of the $c_{u}^{TX}$ coefficient also for the JLEIC1 configuration. To summarize, for the JLEIC dataset we find that the highest sensitivities to the coefficients emanate near the kinematical boundary $Q^{2} = sy$ for low $Q^{2}, y$, and for the smallest and largest value of $E_e$ and $E_p$, respectively.

In regards to the eRHIC results, we begin by consulting Table~\ref{tab:eRHICallbounds}. The first observation we make is that overall, the level of sensitivity of roughly $10^{-5}-10^{-6}$ for the $u$ quark coefficient magnitudes and $10^{-3}-10^{-4}$ for the $d$ quark coefficient magnitudes is comparable with the JLEIC levels, especially for the matching cases of $(E_{e}, E_{p}) = ((5,10), 100)$ GeV as expected. There appears to be a preference for the eRHIC2 configuration, but as with the JLEIC comparison the preference is not significant in the sense that there is not more than an order of magnitude more sensitivity for any coefficient. What is perhaps most interesting about these results is that they further expose and support the feature of increased sensitivity to Lorentz violation at lower $E_e$ for a fixed $E_p$. Additionally, these bounds do not seem to be heavily influenced by an increasing $E_p$.  Furthermore, while the highest sensitivities ultimately come from large $x$, low $Q^{2}$, and large $y$, which is consistent with the JLEIC results, a sizable portion of the best limits come from the contrasting \textit{low} $x$, \textit{large} $y$ regions, with $Q^{2}$ still relatively small; in other words, near the boundary $Q^{2} = sx$. This can be observed from the analogous plots of correlations between these variables for the eRHIC datasets as shown in the lower panels of Fig.~\ref{fig:corr}. This region of sensitivity is consistent with the results of Ref.~\cite{LVdis} which, generally speaking, involved measurements at larger $\sqrt{s}$. Though the bounds that we find here (for the large $x$, low $y$) region are roughly one to two orders of magnitude more sensitive than what was found in Ref.~\cite{LVdis}, the main point is that we have now identified two regions in the kinematical phase space of comparable sensitivity to Lorentz violation: low to moderately low $Q^{2}$ with low $x$, large $y$; or high $x$, low $y$. 

Establishing definitive statements about the patterns in the eRHIC results is not as easily accomplished as compared to the JLEIC results. This is partly due to the greater variability in the eRHIC dataset, coupled with the fact that the systematic errors are calculated differently between the two sets, leading to different patterns in the \textit{correlated} bounds between the two sets. For instance, the pattern of increasing bound sensitivities in $E_{p}$ for a fixed $E_{e}$ is only present for the case of uncorrelated uncertainties and for $E_{e} = 5, 10$ GeV. Since the smallest bounds in this case come from data points that maximize the kinematical preference while minimizing the total uncertainty (which is dominated by systematics and, thus, not strongly reduced by increased statistics), it appears that lower electron energies don't produce competition that changes the trend in more sensitivity to larger $E_{p}$. Even for  $E_{e} = 5, 10$ GeV, with exception for $c_{d}^{\mu\nu}$ with $E_{e} = 5$ GeV, the correlated uncertainties do not follow a pattern of increasing sensitivity with increasing $E_{p}$. As we indicated, some of these features can be understood by comparing the differences in correlated uncertainties (systematics) between the JLEIC and eRHIC datasets.  Taking a closer look at the bounds for $E_{e} = 15, 20$ GeV in Table~\ref{tab:eRHICallbounds} shows that the pattern of increasing sensitivities to larger $E_{p}$ begins to dissolve, even for the case of solely uncorrelated errors. Consulting Fig.~\ref{fig:eRHIC2}, we see that the larger values of $E_{e}$ introduce areas of heightened sensitivity in the low $x$, large $y$ regime, as well as flattening the distribution of bounds overall. Within a given set of fixed $E_{e}$, we can clearly see how increasing $E_{p}$ shifts the distribution of bounds from the low $x$ to high $x$ region, directly exposing the additional dependence on $E_{e}$ and $E_{p}$ individually in the cross section. This feature is not seen in the JLEIC datasets because $E_{e}$ is not large enough to introduce the additional region of low $x$ sensitivity.

The compilation of the best individual and global bounds extracted for both the JLEIC and eRHIC are displayed in Table~\ref{tab:all}. As mentioned, the best individual bound sensitivities for the JLEIC occur for both the $c_{u}^{TX}$ and $c_{u}^{TY}$ coefficients at $(E_{e}, E_{p}) = (5, 100)$ GeV in the region of large $x$, low $Q^{2}$, and low $y$. The global limits, which are extracted by minimizing the combined dataset $\chi^{2}$ distribution, are consistently smaller for all coefficients as expected. Again, there is an even split in which configuration (JLEIC1 or JLEIC2) produces the greatest sensitivities when considering all 12 coefficients. For the eRHIC, the best individual bounds follow roughly the same pattern as the equivalent JLEIC bounds. This indicates, since the colatitudes of the two colliders are very similar, that the effect of the electron beam orientation through the angle $\psi$ is not particularly significant. The JLEIC bounds are found to be slightly more sensitive than those of eRHIC, which we find is predominantly due to the small differences in the supplied $(E_{e}, E_{p}) = (5, 100)$ GeV datasets. 

 Lastly, we return to the issue of potentially finding smaller bounds at lower values of $E_e$ for fixed values of $E_p$, which is what we find for the JLEIC and some of the eRHIC configurations. To illustrate this feature, we consider for example the JLEIC datasets corresponding to $E_p = 100$ GeV and $E_e = 5, 10$ GeV. Speaking purely from the point of view of kinematics, there is an enhancement in the sensitivity to the coefficients for $E_e = 10$ GeV over $E_e = 5$ GeV as we explained above. However, in this case this preference is trumped by the reduction in the relative uncertainty for the equivalent points that yield the best limits. Here, these points correspond to the same values of $x$. Therefore, the lower kinematical cut in $y$ with a lower value of $E_e$ with a fixed $E_p$ will enable a smaller minimum value of $Q^{2}$, which in turn corresponds to a larger value of the cross section as we also explained above. Generally speaking, the larger the value of cross section, the smaller the corresponding statistical uncertainty. This example alone reveals the delicate balance between energetic preferences and experimental uncertainties in the bound extraction procedure.

\subsection{Individual and Global Bounds for Time-Independent Coefficients}
\label{sec:timeind}
In this section we discuss the estimated bounds that can be placed for the six time-independent coefficients $c_{f}^{TT}$, $c_{f}^{TZ}$, and $c_{f}^{ZZ}$ for $f = u$, $d$. We again remind the reader that these coefficients only contribute constant offsets to the leading-order SM result. Here, bounds are extracted by simply finding the data points which minimize the ratio of the square root of the total error to the total numerical factor multiplying the coefficient of interest. Unlike the time-dependent bounds that we discussed in Sec.~\ref{sec:EIC}, these bounds are controlled by the total uncertainty which includes theoretical (mainly emanating from PDFs and evaluated following Ref.~\cite{Owens:2012bv}), statistical and experimental systematic contributions. We present the expected best individual and global bounds for both configurations of the JLEIC and eRHIC in Table~\ref{tab:timeindependent}. 

Overall, the bounds on these coefficients are of a similar magnitude to the time-dependent bounds---see, e.g.,  Table~\ref{tab:all}. An interesting feature emerges from the coefficients $c_{f}^{ZZ}$ in particular---it is observed that these bounds for the JLEIC case vary by roughly an order of magnitude between JLEIC1/JLEIC2, whereas the equivalent eRHIC bounds do not. This is due to the fact that, under the rotation given in Eq.~\eqref{eq:rotation}, the coefficients $c_{f}^{ZZ}$ inherit a factor proportional to $\cos(2\psi)$. It turns out that the JLEIC1 configuration with $\psi \simeq \ang{47.6}$ is the only one of the four configurations which is near the ``least optimum" angle possible, which in turn generates the largest bound. In any case, we caution the reader in the interpretations of these bounds. Unlike in the case of the coefficients which generate sidereal time variation in the cross section (something that no mechanism in the SM can do), constant shifts in the conventional SM result could be argued to emanate from a number of factors. What may be most pressing of an issue here is the fact that, if these effects of Lorentz violation are indeed present, they may already be contained within, e.g., the PDFs which we have used to extract the bounds. If so, this would generate an inconsistency in the bound extraction procedure. Since it is currently unknown whether or not this is the case, we cannot ascribe a reliably meaningful result to these bounds. We leave a more in depth study of these details for a future work.
\begin{table}[h]
\begin{tabular}{|lccccccc|}
\hline
& {\bf JLEIC} 
&\;\;& {\bf JLEIC} 
&\;\;&  {\bf eRHIC} 
&\;\;&  {\bf eRHIC} 
 \\
&  {\bf Individual} 
&\;\;& {\bf Global} 
&\;\;&  {\bf Individual} 
&\;\;&  {\bf Global} 
 \\[0.1cm] \hline \\[-0.3cm]
$|c_{u}^{TT}|$ & 0.5 && 0.3 && 0.6 && 0.3\\ 
               & 0.5 && 0.3 && 0.5 && 0.3\\ 
$|c_{u}^{TZ}|$ & 0.5 && 0.3 && 0.5 && 0.3\\ 
               & 0.7  && 0.4 && 2  && 1\\ 
$|c_{u}^{ZZ}|$ & 30 && 20 && 2   && 1\\
               & 3 && 2 && 2   && 1
\\[0.1cm] \hline \\[-0.3cm]
$|c_{d}^{TT}|$ & 8 && 4 && 10  && 5\\ 
               & 8 && 4 && 10 && 4\\ 
$|c_{d}^{TZ}|$ & 9 && 5 && 10 && 4\\ 
               & 10 && 6 && 30 && 10\\ 
$|c_{d}^{ZZ}|$ &  500 && 300 && 40 && 20\\
               & 60 && 30&& 30 && 10\\ 
\hline
\end{tabular}
\caption{Expected time-independent best individual and global bounds for the JLEIC and eRHIC. All bounds are given in units of $10^{-5}$. For each coefficient magnitude, we give the bounds for both electron beam orientations (see the caption in Tables.~\ref{tab:JLEICallbounds} and \ref{tab:eRHICallbounds} for further information). \label{tab:timeindependent}}
\end{table}

\section{Conclusions}
In this work we have explored the potential constraints on Lorentz-violating unpolarized $e$-$p$ DIS with the two currently proposed EIC designs. Our results indicate that both the JLEIC and eRHIC can offer increased sensitivities to the coefficients for Lorentz violation which induce a sidereal time variation in the scattering cross section by revealing a new kinematical regime in which these effects, as encapsulated by Eq.~\eqref{eq:photonex}, are enhanced by roughly one to two orders of magnitude over previous estimates \cite{LVdis} which focused on HERA data. We also provided predictions for the six coefficients which contribute as constant offsets to the leading-order SM cross section.

In light of our results, it is reasonable to suggest that the EIC can be a useful tool for studying deviations from exact Lorentz symmetry in unpolarized $e$-$p$ DIS and related processes. Future studies focusing on the large $x$, low $Q^{2}$, and large $y$ region of the phase space at low electron energies with more refined $(x,Q^{2})$ binning could be performed to give a more realistic idea of what bounds could be achieved in an actual experiment where the time stamps of the events culminating in a measurement are known. Though we have checked that increasing the number of sidereal time bins does not yield a substantial improvement in regards to the extracted limits, it may be the case that a larger number of bins might be required to wash out potentially sizable day/night effects (which are periodic with period $T = 24 \;{\rm hr}$). However, the size of these effects can only be estimated once the collider is operational. Additionally, given that one of the main motivations behind the EIC is to explore polarization effects in the structure of hadrons, one could conceivably investigate the prospects for testing Lorentz symmetry in related processes such as polarized DIS. In any event, we view this work as further support that that the EIC will be a promising and important tool for searches for new physics beyond the standard model.

\section*{ACKNOWLEDGMENTS}
We extend thanks to V. A. Kosteleck\'y for many useful discussions. Simulated data for the JLEIC and eRHIC collider were generated and supplied by A. Accardi (JLab/Hampton University) and Y. Furletova (JLab), and by E. C. Aschenauer (BNL) and B. Page (BNL), respectively. This work was supported in part by the United States Department of Energy under grants DE-AC05-06OR23177, DE-FG0287ER40365, DE-SC0010120, and by the Indiana University Center for Spacetime Symmetries.

\bibliographystyle{apsrev4-1}
\bibliography{EIC}

\newpage

\begin{table}[t]
\begin{tabular}{|lccccccccc|}
\hline
{\bf JLEIC}&  {\bf (10, 20)} 
&\;\;& {\bf (10, 60)} 
&\;\;&  {\bf (10, 80)} 
&\;\;&  {\bf (10, 100)} 
&\;\;&  {\bf (5, 100)} 
\\[0.1cm] \hline \\[-0.3cm]
$|c_{u}^{TX}|$ & $1\;[8]$ && $0.8\;[4]$ && $0.7\;[1]$ && $0.5\;[0.9]$ && 0.04\;[0.2]\\
 & $1\; [7]$ && $0.7\; [4]$&& $0.6\; [1]$&& $0.4\; [0.8]$ && 0.04\;[0.2]\\
 
$|c_{u}^{TY}|$ & $1\;[8]$ && $0.8\;[4]$ && $0.7\;[1]$ && $0.5\;[0.9]$ && 0.04\;[0.2]\\
 & $1\; [7]$ && $0.7\; [4]$&& $0.6\; [1]$&& $0.4\; [0.8]$ && 0.04\;[0.2]\\
 
$|c_{u}^{XZ}|$ &$2\;[10]$ && $1\;[6]$ && $1\;[2]$ && $0.8\;[2]$ && 0.07\;[0.4]\\
& $2\; [20]$ && $2\; [8]$&& $1\; [2]$&& $1\; [2]$ && 0.08\;[0.4]\\

$|c_{u}^{YZ}|$ & $2\;[10]$ && $1\;[6]$ && $1\;[2]$ && $0.8\;[2]$ && 0.07\;[0.4]\\
& $2\; [20]$ && $2\; [8]$&& $1\; [2]$&& $1\; [2]$ && 0.09\;[0.4]\\

$|c_{u}^{XY}|$ & $7\;[50]$ && $5\;[20]$ && $4\;[7]$ && $3\;[6]$ && 0.3\;[1]\\
& $3\; [20]$ && $2\; [10]$&& $2\; [3]$&& $1\; [3]$ && 0.1\;[0.7]\\

$|c_{u}^{XX} - c_{u}^{YY}|$ & $6\;[40]$ && $4\;[20]$ && $4\;[6]$ && $3\;[5]$ && 0.2\;[1]\\
& $7\; [50]$ && $4\; [20]$&& $4\; [6]$&& $3\; [5]$ && 0.2\;[1]\\
[0.1cm] \hline \\[-0.3cm]

$|c_{d}^{TX}|$ & $20\;[100]$ && $7\;[60]$ && $6\;[20]$ && $6\;[10]$ && 0.7\;[4]\\
& $10\; [100]$ && $6\; [60]$&& $5\; [20]$&& $5\; [10]$ && 0.7\;[3]\\

$|c_{d}^{TY}|$ & $20\;[100]$ && $7\;[70]$ && $6\;[20]$ && $6\;[10]$ && 0.7\;[4]\\
& $10\; [100]$ && $6\; [60]$&& $5\; [20]$&& $5\; [10]$ && 0.6\;[3]\\

$|c_{d}^{XZ}|$ & $20\;[200]$ && $10\;[100]$ && $9\;[30]$ && $9\;[20]$ && 1\;[6]\\
& $30\; [300]$ && $10\; [100]$&& $10\; [40]$&& $10\; [30]$ && 1\;[7]\\

$|c_{d}^{YZ}|$ & $20\;[200]$ && $10\;[100]$ && $9\;[30]$ && $9\;[20]$ && 1\;[6]\\
& $30\; [300]$ && $10\; [100]$&& $10\; [40]$&& $10\; [30]$ && 1\;[8]\\

$|c_{d}^{XY}|$ &  $90\;[900]$ &&  $40\;[400]$ && $40\;[100]$ &&$40\;[90]$ && 5\;[20]\\
& $40\; [400]$ && $20\; [200]$&& $20\; [50]$&& $20\; [40]$ && 2\;[10]\\

$|c_{d}^{XX} - c_{d}^{YY}|$ & $80\;[700]$ && $40\;[400]$ && $30\;[90]$ && $30\;[70]$ && 4\;[20]\\
& $80\; [800]$ && $40\; [400]$&& $30\; [100]$&& $30\; [70]$ && 4\;[20]\\
\hline
\end{tabular}
\caption{ Summary of all expected bounds for the JLEIC given in units of $10^{-5}$. The five columns denote the laboratory frame electron and proton energies \tb{$(E_{e}, E_{p})$} in GeV. The bounds with brackets correspond to the case of uncorrelated systematic uncertainties between binned data, and the bounds without brackets correspond to assuming 100\% correlation between systematic uncertainties. For each coefficient magnitude, we give the bounds for both electron beam orientations: $\psi = \ang{47.6}$ followed by $\psi =\ang{-35.0}$ NE.\label{tab:JLEICallbounds}}
\end{table}

\begin{table}[t]
\begin{tabular}{|lccccccccccc|}
\hline
{\bf eRHIC}&  {\bf (5,50)}
&\;\;& {\bf [5,100]} 
&\;\;&  {\bf [5,250]} 
&\;\;&  {\bf [10,50]} 
&\;\;&  {\bf [10,100]} 
&\;\;&  {\bf [10,250]}  \\[0.1cm] \hline \\[-0.3cm]
$|c_{u}^{TX}|$ & 0.1\; [0.9] && 0.2\; [0.8] && 0.1\; [0.5] && 0.6\; [3] && 0.6\; [2] && 0.4\; [0.9] \\ 
               & 0.09\; [0.6] && 0.1\; [0.5] && 0.09\; [0.3] && 0.4\; [2] && 0.4\; [1] && 0.3\; [0.6] \\ 
$|c_{u}^{TY}|$ & 0.1\; [0.9] && 0.2\; [0.7] && 0.1\; [0.5] && 0.6\; [3] && 0.6\; [2] && 0.4\; [0.9] \\ 
               & 0.09\; [0.6] && 0.1\; [0.5] && 0.09\; [0.3] && 0.4\; [2] && 0.4\; [1] && 0.3\; [0.6] \\ 
$|c_{u}^{XZ}|$ & 0.2\; [1] && 0.2\; [1] && 0.2\; [0.7] && 0.8\; [4] && 0.9\; [2] && 0.5\; [1] \\ 
               & 0.4\; [3] && 0.5\; [2] && 0.4\; [2] && 2\; [9] && 2\; [5] && 1\; [3] \\ 
$|c_{u}^{YZ}|$ & 0.2\; [1] && 0.2\; [1] && 0.2\; [0.7] && 0.9\; [4] && 0.9\; [2] && 0.5\; [1] \\ 
               & 0.4\; [3] && 0.5\; [2] && 0.4\; [2] && 2\; [9] && 2\; [5] && 1\; [3] \\ 
$|c_{u}^{XY}|$ & 0.5\; [3] && 0.6\; [3] && 0.5\; [2] && 2\; [10] && 2\; [6] && 1\; [3] \\ 
               & 0.2\; [1] && 0.2\; [1] && 0.2\; [0.7] && 1\; [4] && 1\; [3] && 0.6\; [1] \\ 
$|c_{u}^{XX}-c_{u}^{YY}|$ &1\; [9] && 2\; [8] && 1\; [5] && 6\; [30] && 7\; [20] && 4\; [9] \\ 
                          & 0.9\; [7] && 1\; [5] && 1\; [4] && 5\; [20] && 5\; [10] && 3\; [7] \\
[0.1cm] \hline \\[-0.3cm]
$|c_{d}^{TX}|$ & 2\; [20] && 2\; [10] && 1\; [10] && 6\; [60] && 6\; [30] && 4\; [20] \\ 
               & 1\; [10] && 1\; [10] && 0.9\; [6] && 4\; [40] && 4\; [20] && 2\; [10] \\ 
$|c_{d}^{TY}|$ & 2\; [20] && 2\; [10] && 1\; [10] && 6\; [60] && 6\; [30] && 4\; [20] \\ 
               & 1\; [10] && 1\; [10] && 0.9\; [7] && 4\; [40] && 4\; [20] && 2\; [10] \\ 
$|c_{d}^{XZ}|$ & 3\; [20] && 2\; [20] && 2\; [10] && 7\; [70] && 8\; [40] && 5\; [20] \\ 
               & 6\; [60] && 5\; [50] && 4\; [30] && 20\; [200] && 20\; [100] && 10\; [50] \\ 
$|c_{d}^{YZ}|$ & 3\; [20] && 2\; [20] && 2\; [10] && 7\; [70] && 8\; [40] && 5\; [20] \\ 
               & 6\; [60] && 5\; [50] && 4\; [30] && 20\; [200] && 20\; [100] && 10\; [50] \\ 
$|c_{d}^{XY}|$ & 7\; [70] && 6\; [50] && 5\; [30] && 20\; [200] && 20\; [100] && 10\; [60] \\ 
               & 3\; [30] && 3\; [20] && 2\; [10] && 8\; [90] && 9\; [50] && 5\; [30] \\ 
$|c_{d}^{XX}-c_{d}^{YY}|$ &20\; [200] && 20\; [200] && 10\; [100] && 60\; [600] && 70\; [400] && 40\; [200] \\ 
                          & 10\; [100] && 10\; [100] && 9\; [70] && 40\; [400] && 50\; [200] && 30\; [100] \\
\hline
\hline
\vphantom{$\Big($}&  {\bf [15,50]} 
&\;\;& {\bf [15,100]} 
&\;\;&  {\bf [15,250]} 
&\;\;&  {\bf [20,50]} 
&\;\;&  {\bf [20,100]} 
&\;\;&  {\bf [20,250]}  \\[0.1cm] \hline \\[-0.3cm]
$|c_{u}^{TX}|$ & 1\; [2] && 1\; [3] && 0.8\; [2] && 1\; [6] && 1\; [4] && 1\; [2] \\ 
               & 0.7\; [2] && 0.8\; [2] && 0.5\; [1] && 0.8\; [4] && 0.8\; [3] && 0.8\; [1] \\ 
$|c_{u}^{TY}|$ & 1\; [2] && 1\; [3] && 0.8\; [2] && 1\; [7] && 1\; [4] && 1\; [2] \\ 
               & 0.7\; [2] && 0.7\; [2] && 0.5\; [1] && 0.8\; [4] && 0.8\; [3] && 0.8\; [1] \\ 
$|c_{u}^{XZ}|$ & 1\; [3] && 1\; [4] && 1\; [2] && 2\; [9] && 2\; [6] && 2\; [3] \\ 
               & 3\; [7] && 3\; [10] && 3\; [5] && 4\; [20] && 4\; [10] && 4\; [6] \\ 
$|c_{u}^{YZ}|$ & 1\; [3] && 1\; [4] && 1\; [2] && 2\; [9] && 2\; [6] && 2\; [3] \\ 
               & 3\; [7] && 3\; [10] && 2\; [5] && 4\; [20] && 4\; [10] && 4\; [6] \\ 
$|c_{u}^{XY}|$ & 3\; [8] && 4\; [10] && 3\; [6] && 4\; [20] && 4\; [20] && 4\; [8] \\ 
               & 1\; [3] && 2\; [5] && 1\; [3] && 2\; [10] && 2\; [7] && 2\; [3] \\ 
$|c_{u}^{XX}-c_{u}^{YY}|$ &10\; [20] && 10\; [30] && 8\; [20] && 10\; [70] && 10\; [50] && 10\; [20] \\ 
                          & 7\; [20] && 8\; [20] && 6\; [10] && 9\; [50] && 9\; [30] && 9\; [20] \\
[0.1cm] \hline \\[-0.3cm]
$|c_{d}^{TX}|$ & 4\; [10] && 10\; [20] && 7\; [30] && 5\; [40] && 5\; [60] && 10\; [40] \\ 
               & 3\; [7] && 7\; [10] && 5\; [20] && 3\; [20] && 3\; [40] && 8\; [30] \\ 
$|c_{d}^{TY}|$ & 4\; [9] && 10\; [20] && 7\; [30] && 5\; [40] && 5\; [60] && 10\; [40] \\ 
               & 3\; [7] && 7\; [10] && 5\; [20] && 3\; [20] && 3\; [40] && 8\; [30] \\ 
$|c_{d}^{XZ}|$ & 5\; [10] && 10\; [30] && 10\; [40] && 6\; [50] && 6\; [80] && 20\; [50] \\ 
               & 10\; [30] && 30\; [60] && 20\; [100] && 20\; [100] && 10\; [200] && 40\; [100] \\ 
$|c_{d}^{YZ}|$ & 5\; [10] && 10\; [30] && 10\; [50] && 6\; [50] && 6\; [80] && 10\; [50] \\ 
               & 10\; [30] && 30\; [60] && 20\; [100] && 20\; [100] && 10\; [200] && 30\; [100] \\ 
$|c_{d}^{XY}|$ & 10\; [30] && 40\; [70] && 30\; [100] && 20\; [100] && 20\; [200] && 40\; [100] \\ 
               & 6\; [10] && 20\; [30] && 10\; [50] && 7\; [50] && 7\; [90] && 20\; [60] \\ 
$|c_{d}^{XX}-c_{d}^{YY}|$ &40\; [100] && 100\; [200] && 80\; [300] && 50\; [400] && 50\; [600] && 100\; [400] \\ 
                          & 30\; [70] && 80\; [100] && 50\; [200] && 30\; [300] && 30\; [400] && 80\; [300] \\
\hline
\end{tabular}
\caption{Summary of all expected bounds for the eRHIC.  Bounds for the electron beam orientations $\psi = \ang{-78.5}$ followed by $\psi = \ang{-16.8}$ NE are shown. See the caption in Table~\ref{tab:JLEICallbounds} for further details.\label{tab:eRHICallbounds}}
\end{table}

\end{document}